\documentclass[12pt]{article}

\usepackage{graphicx}

\usepackage{latexsym} 
\usepackage{amssymb}  

\newcommand{\Dash}{\boldmath $-$}
\newcommand\eqn[1]{(\ref{#1})}      
\newcommand\Eqn[1]{Eq.~(\ref{#1})}  

\newcommand{\de}{\delta}
\newcommand{\De}{\Delta}

\begin{document}

\title{{\bf Dense quark matter in nature}}

\author{Mark Alford \\[1ex]
Physics Department \\ Washington University CB 1105 \\
Saint Louis, MO 63130 \\ USA
}

\date{Dec 1, 2003}

\begin{titlepage}
\maketitle

\begin{abstract}
According to quantum chromodynamics (QCD),
matter at ultra-high densities will take the form
of a color-superconducting quark liquid, in which there is
a condensate of Cooper pairs of quarks near the Fermi surface.
I present a review of the physics of color superconductivity.
I give particular attention to the recently proposed gapless CFL 
(gCFL) phase, which has unusual properties such as
quasiquarks with a near-quadratic dispersion relation,
and which may well be the favored phase of quark matter in the
density range relevant to compact stars.
I also discuss
the effects of color superconductivity on the
mass-radius relationship of compact stars, showing that
one would have to fix the bag constant by other measurements
in order to see the effects of color superconductivity.
An additional parameter in the quark matter equation of state
connected with perturbative corrections allows quark matter
to imitate nuclear matter over the relevant density range,
so that hybrid stars
can show a mass-radius relationship very similar to that
of nuclear matter, and their masses can reach $1.9~M_\odot$.
\end{abstract}

\end{titlepage}

\section{Introduction}
\label{sec:intro}
One of the most striking features of QCD is asymptotic freedom: the
force between quarks becomes arbitrarily weak as the characteristic
momentum scale of their interaction grows larger.  This immediately
suggests that at sufficiently high densities and low temperatures,
matter will consist of a Fermi sea of essentially free quarks,
whose behavior is dominated by  the
high-momentum quarks that live at the Fermi surface.

However, over the last few years it has become clear
that the phase diagram
of QCD is much richer than this.
In addition to the hadronic phase with which we are
familiar and the quark gluon plasma (QGP) that is predicted
to lie at temperatures above $170~{\rm MeV}$, there is a
whole family of ``color superconducting'' phases, that are expected
to occur at high density and low temperature \cite{Reviews}.
The essence of color superconductivity is quark pairing, driven by
the BCS mechanism, which operates when there exists
an attractive interaction between fermions at a Fermi surface.
The QCD quark-quark interaction is strong, and is attractive
in many channels, so we expect cold dense quark matter to {\em generically}
exhibit color superconductivity.
Moreover, quarks, unlike electrons, have color and flavor as well as spin
degrees of freedom, so many different patterns of pairing are possible.
This leads us to expect a rich phase structure
in matter beyond nuclear density.

Color superconducting quark matter may occur naturally in the
universe, in the cold dense cores of compact (``neutron'') stars,
where densities are above nuclear density, and temperatures are of the
order of tens of {\rm keV}.  (It might conceivably be possible to
create it in future low-energy heavy ion colliders, such as the Japan
Proton Accelerator Research Complex (J-PARC) or the Compressed
Baryonic Matter facility at GSI Darmstadt.)  Up to now, most work on
signatures has focussed on properties of color superconducting quark
matter that would affect observable features of compact stars, and I
will discuss some of these below.

\section{Phase diagram of quark matter}

\begin{figure}[t]
\parbox{0.48\hsize}{
\begin{center} Light strange quark\\ or large pairing gap \end{center}
 \includegraphics[width=\hsize]{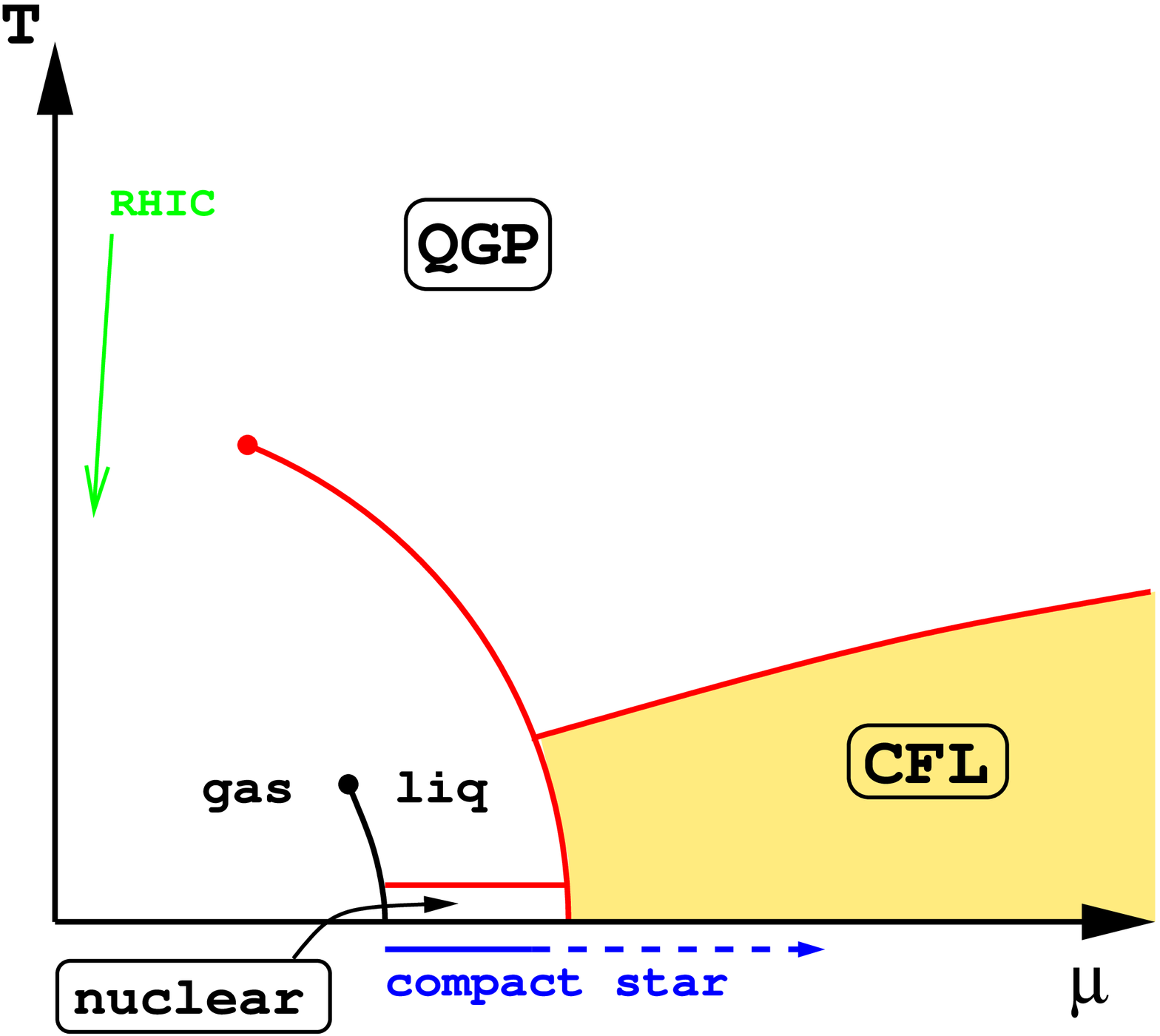}
}
\parbox{0.48\hsize}{
\begin{center} Heavy strange quark\\ or small pairing gap \end{center}
 \includegraphics[width=\hsize]{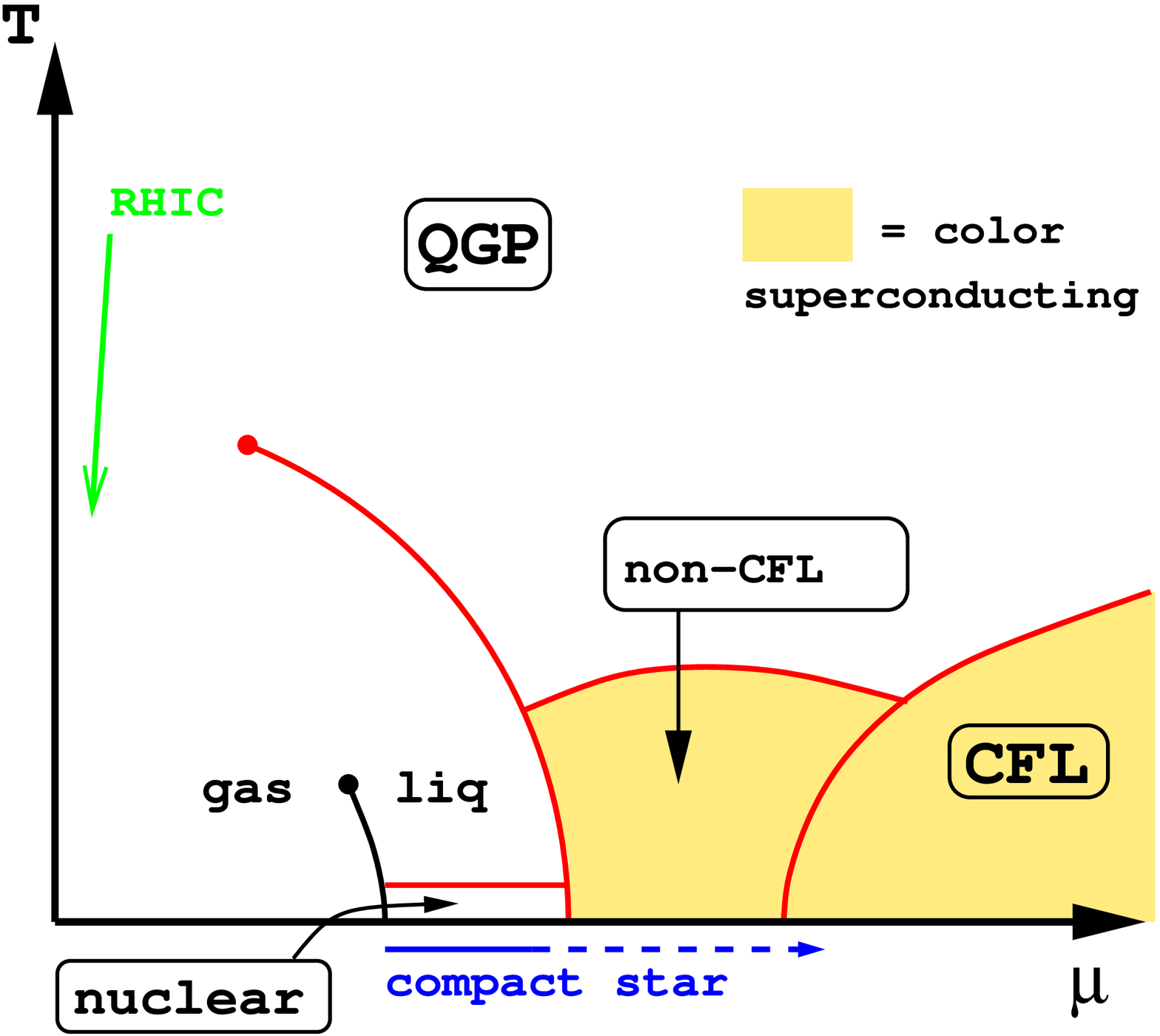}
}
\caption{Conjectured phase diagrams for QCD in the real world.
For small $M_s^2/\De$ there is a direct transition from
nuclear matter to color-flavor locked color
superconducting quark matter. For large  $M_s^2/\De$
there is an intermediate phase where the strange quark
pairs in some other way. Depending on the strength of instanton 
interactions, the CFL phase may include $K^0$ condensation.}
\label{fig:phase}
\end{figure}

In the real world there are two light quark flavors, the up 
($u$) and down ($d$), with
masses $\lesssim 10~{\rm MeV}$, and a medium-weight flavor, the strange 
($s$) quark, with mass $\sim 150~{\rm MeV}$.
The strange quark therefore plays a crucial role in the phases of QCD,
and we expect it to remain fully paired with the light flavors as
long as $\mu \gg M_s^2/\De$, where $\De$ is a gap parameter for
the pairing of the strange quark.
Fig.~\ref{fig:phase} shows two conjectured phase diagrams for QCD.
One panel is for small $M_s^2/\De$, in which case the strange quark's
mass never breaks its pairing with the light flavors, so
there is a direct transition
from nuclear matter to color-flavor-locked (CFL) quark matter \cite{ARW3}. 
In the CFL phase the strange
quark participates symmetrically with the up and down quarks
in Cooper pairing---this is described
in more detail in section \ref{sec:CFL}.
The other panel in Fig.~\ref{fig:phase} is for large $M_s^2/\De$, 
in which case  the strange quark is too heavy to pair
symmetrically with the light quarks at medium densities, 
then there will be an interval of some other phase or phases.
These may well include the recently proposed gapless CFL
phase \cite{gCFL}, which will be described in section
\ref{sec:gCFL}, although other possibilities such as
crystalline color superconductivity \cite{OurLOFF} or
some form of single-flavor pairing \cite{oneflav,IwaIwa}
have been suggested.


\section{Review of color superconductivity}

The essential physics of color superconductivity is the same
as that underlying conventional superconductivity in metals
\cite{oldcolorSC,IwaIwa,newcolorSC}.
As mentioned above, asymptotic freedom of QCD means that
at sufficiently high density and low temperature,
there is a  Fermi surface of almost free quarks. 
The interactions between
quarks near the Fermi surface are certainly attractive in some channels
(quarks bind together to form baryons)
and it was shown by Bardeen, Cooper, and
Schrieffer (BCS) \cite{BCS} that if there is {\em any} channel in which the
interaction is attractive, then there is a state
of lower free energy than a simple Fermi surface. That state arises
from  a complicated coherent 
superposition of pairs of particles (and holes)---``Cooper pairs''.

Attractive interactions play a crucial role in the BCS mechanism
for the formation of Cooper pairs. This can easily be understood
in an intuitive way.
The Helmholtz free energy is $F= E-\mu N$, where $E$ is
the total energy of the system, $\mu$ is the chemical potential, and
$N$ is the number of fermions. The Fermi surface is defined by a
Fermi energy $E_F=\mu$, at which the free energy is minimized, so
adding or subtracting a single particle costs zero free energy. 
Now switch on a weak attractive interaction.
It costs no free energy to
add a pair of particles (or holes), and if they have the
right quantum numbers then the attractive
interaction between them will lower the free energy of the system.
Many such pairs will therefore
be created in the modes near the Fermi surface, and these pairs,
being bosonic, will form a condensate. The ground state will be a
superposition of states with all numbers of pairs, breaking the
fermion number symmetry. 

Since pairs of quarks cannot be color singlets,
the resulting condensate will break the local color symmetry
$SU(3)_{\rm color}$.  We call this ``color superconductivity''.
Note that the quark pairs play the same role here as the Higgs particle
does in the standard model: the color-superconducting phase
can be thought of as the Higgs phase of QCD.

\subsection{Three flavors: Color-flavor locking (CFL)}
\label{sec:CFL}

The favored pairing pattern at high densities, where
the strange quark Fermi momentum is close to the up and down
quark Fermi momenta, is ``color-flavor locking'' (CFL) \cite{ARW3}.
This has been confirmed by both NJL \cite{ARW3,SW-cont,HsuCFL} and 
gluon-mediated interaction calculations \cite{3flavpert}.
The CFL pairing pattern is
\begin{equation}
\begin{array}{c}
\langle q^\alpha_i q^\beta_j \rangle^{\phantom\dagger}_{1PI}
\propto C \gamma_5 \Bigl(
 (\kappa+1)\delta^\alpha_i\delta^\beta_j + (\kappa-1) \delta^\alpha_j\delta^\beta_i \Bigr)
  \\[2ex]
 {[SU(3)_{\rm color}]}
 \times \underbrace{SU(3)_L \times SU(3)_R}_{\displaystyle\supset [U(1)_Q]}
 \times U(1)_B 
 \to \underbrace{SU(3)_{C+L+R}}_{\displaystyle\supset [U(1)_{{\tilde Q} }]} 
  \times \mathbb{Z}_2
\end{array}
\label{CFLcond}
\end{equation}
Color indices $\alpha,\beta$ and flavor indices $i,j$ run from 1 to 3,
Dirac indices are suppressed,
and $C$ is the Dirac charge-conjugation matrix.
The term multiplied by $\kappa$ corresponds to pairing in the
$({\bf 6}_S,{\bf 6}_S)$, which
although not energetically favored
breaks no additional symmetries and so
$\kappa$ is in general small but not zero 
\cite{ARW3,3flavpert,PisarskiCFL}.
The Kronecker deltas connect
color indices with flavor indices, so that the condensate is not
invariant under color rotations, nor under flavor rotations,
but only under simultaneous, equal and opposite, color and flavor
rotations. Since color is only a vector symmetry, this
condensate is only invariant under vector flavor+color rotations, and
breaks chiral symmetry. The features of the CFL pattern of condensation are
\begin{itemize}
\setlength{\itemsep}{-0.7\parsep}
\item[\Dash] The color gauge group is completely broken. All eight gluons
become massive. This ensures that there are no infrared divergences
associated with gluon propagators.
\item[\Dash]
All the quark modes are gapped. The nine quasiquarks 
(three colors times three flavors) fall into an ${\bf 8} \oplus {\bf 1}$
of the unbroken global $SU(3)$, so there are two
gap parameters. The singlet has a larger gap than the octet.
\item[\Dash] 
A rotated electromagnetism (``${\tilde Q} $'')
survives unbroken. It is a combination
of the original photon and one of the gluons.
\item[\Dash] Two global symmetries are broken,
the chiral symmetry and baryon number, so there are two 
gauge-invariant order parameters
that distinguish the CFL phase from the QGP,
and corresponding Goldstone bosons which are long-wavelength
disturbances of the order parameter. 
When the light quark mass is non-zero it explicitly breaks
the chiral symmetry and gives a mass
to the chiral Goldstone octet, but the CFL phase is still
a superfluid, distinguished by its baryon number breaking.
\item[\Dash]
The symmetries of the
3-flavor CFL phase are the same as those one might expect for 3-flavor
hypernuclear matter \cite{SW-cont}, so it is possible that there is
no phase transition between them.
\end{itemize}

In a real compact star we must require electromagnetic and color
neutrality \cite{BaymIida,AR-02} (possibly
via mixing of oppositely-charged phases),
allow for equilibration under the weak interaction, and include a
realistic mass for the strange quark.  These factors tend to pull
apart the Fermi momenta of the different quark species, imposing an
energy cost on cross-species pairing.

The requirement of neutrality
penalizes the 2SC phase relative to the
CFL phase. This can be shown by analyzing a generic expansion
of the free energy in powers of $m_s/\mu$ \cite{AR-02}
or by an NJL calculation \cite{Steiner:2002gx} that
handles $m_s\sim\mu$ and includes the
coupling between the chiral condensate and quark condensate gap
equations.  The net result, assuming that
mixed phases are excluded by the surface energy cost \cite{ARRW}
(see Section.~\ref{sec:phenomenology}), is that there
is no (or very little) density range in which 2SC is the phase with
the lowest free energy: unpaired or CFL-paired quark matter are
generally favored over 2SC.

\subsection{Gapless CFL (gCFL)}
\label{sec:gCFL}

Introducing a strange quark mass leads to new color-superconducting
phases.
The dimensionless parameter that expresses the effect of the 
strange quark mass is $M_s^2/(\mu\De)$, which tells us how 
the pairing gap $\De$ of the strange quark compares with the
amount by which the strange quark mass is trying to separate
the strange quark Fermi surface from those of the light quarks.
In reality, both $M_s$ and $\De$ will depend on $\mu$, but
in the rest of this section we will treat $M_s$ as a
parameter, and discuss the effects of
varying $M_s^2/\mu$, invoking model-independent arguments
as well as results from a NJL model in which the pairing
gap at $\mu=500$~MeV and $M_s=0$ is $\De_0=25$~MeV.

At very high densities the strange quark mass is small relative
to the chemical potential ($M_s^2/\mu \ll \De_{CFL}$)
and it may, depending on the size of
instanton effects \cite{Schaefer:2002gf}, induce a flavor rotation of
the CFL condensate known as ``kaon condensation'' \cite{BedaqueSchaefer},
which breaks isospin.  As one reduces the density 
(increasing $M_s^2/\mu$) the strange quark mass
becomes more important, and it now seems \cite{gCFL} that there is a
smooth transition into a gapless CFL
phase ``gCFL'' at $M_s^2/\mu=2\De_{CFL}$. This will be discussed
in more detail below.
At lower densities things become complicated. 
The strange quark mass
and the requirements of color and electric neutrality impose
a free energy cost for
keeping the Fermi momenta of different flavors locked together
so that they can pair with each other, and
when $M_s^2/\mu$ is large this cost is too great to be compensated
by the resultant pairing energy.
One might expect the strange quark to decouple first, leading
to ``2SC'' pairing between up and down only
\cite{SW-cont,ABR2+1}, but neutrality constraints disfavor this \cite{AR-02}.
However,  we do not expect the
favored phase in this region to be rigorously unpaired.
There may be a range of $M_s^2/\mu$ in which the different flavors
are able to maintain some pairing by switching from BCS pairing
to ``LOFF'' crystalline pairing, involving only part
of the Fermi surface \cite{LOFF,OurLOFF,Bowers:2002xr} 
(see Section~\ref{sec:phenomenology}). 
Another possibility is that each flavor
simply pairs with itself \cite{oneflav,IwaIwa}.
At some point that cannot be predicted with current techniques,
there will be a transition from quark matter to baryonic matter.

The gapless CFL phase, then, is the ``medium-high density'' phase
of QCD. It  involves a more complicated pairing pattern than
\eqn{CFLcond},
\begin{equation}
\langle \psi^\alpha_a C\gamma_5 \psi^\beta_b \rangle \sim 
\Delta_1 \epsilon^{\alpha\beta 1}\epsilon_{ab1} + 
\Delta_2 \epsilon^{\alpha\beta 2}\epsilon_{ab2} + 
\Delta_3 \epsilon^{\alpha\beta 3}\epsilon_{ab3}\ .
\label{gCFLcond}
\end{equation} 
Here $\psi^\alpha_a$ is a quark of color $\alpha=(r,g,b)$ 
and flavor $a=(u,d,s)$;
the condensate is a Lorentz scalar, antisymmetric in Dirac indices,
antisymmetric in color 
(the channel with the strongest
attraction between quarks), and consequently
antisymmetric in flavor. 
The $\De_i$ parameterize the pairing as follows:
\begin{itemize}\setlength{\itemsep}{0ex} 
\item 
 The $rd$ and $gu$ quarks pair with gap parameter $\De_3$.
\item The $bu$ and $rs$ quarks pair with gap parameter $\De_2$.
\item
The $gs$ and $bd$ quarks pair with gap parameter $\De_1$.
\item
The $ru$, $gd$ and $bs$ quarks pair among each other in a fashion
involving all three gap parameters $\De_1$, $\De_2$ and $\De_3$.
\end{itemize}

\begin{figure}[htb]
\begin{center}
\includegraphics[width=0.7\textwidth]{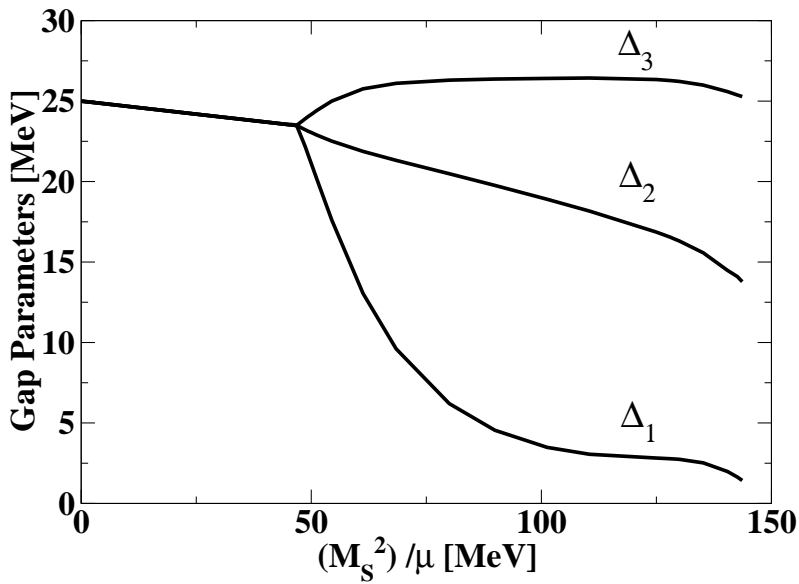}
\end{center}
\vspace{-0.25in}
\caption{
Gap parameters $\De_3$, $\De_2$, and $\De_1$
as a function of $M_s^2/\mu$ for $\mu=500$~MeV, in an NJL model
where $\De_0=25$~MeV \cite{gCFL}.
There is a second order phase
transition between the CFL phase 
and the gapless CFL phase at
$M_s^2/\mu = 2 \De$.
}
\label{fig:gaps}
\end{figure}

As with CFL pairing, there is a ``rotated electromagnetism''
generated by a mixture $\tilde Q$ of the photon and a gluon.
However, unlike CFL, which is a $\tilde Q$ insulator,
gCFL 
has some $\tilde Q$-charged gapless modes
(assuming we include electrons as well as quark matter), and is therefore
a $\tilde Q$ conductor. At the continuous phase transition between
CFL and gCFL, there is a simultaneous partial unpairing of
the $gs$ and $bd$ quarks, and the $bu$ and $rs$ quarks
(see Fig.~\ref{fig:gaps}).
The corresponding gaps $\De_1$ and $\De_2$ drop 
($\De_1$ much more rapidly) as
$M_s^2/\mu$ increases, while the gap parameter $\De_3$ for the
$rd$ and $gu$ quarks, which remain robustly paired, rises.

The difference in behavior of $\De_1$, $\De_2$ and $\De_3$ reflects
the ways in which different quarks experience the stress of an
increasing $M_s$, and the requirement of neutrality.
To understand this better, let us
study their dispersion relations.

For a pair of massless quarks that pair with gap parameter $\De$,
the dispersion relation is
\begin{equation}
E(p) = \Bigl| \de\mu \pm \sqrt{ (p-\bar \mu)^2 + \De^2} \Bigr|
\label{disprel}
\end{equation}
where the individual chemical potentials of the quarks
are $\bar\mu \pm \de \mu$. As long as the chemical potentials
pulling the two species apart are not too strong,
Cooper pairing occurs in all modes
\begin{equation}
\mbox{pairing criterion:}\quad |\de\mu|<\De
\label{pairing}
\end{equation}
However when this condition is violated
there are gapless ($E=0$) modes at momenta
\begin{equation}
p_{\rm gapless} = \bar\mu \pm \sqrt{\de\mu^2-\De^2}
\label{blocking}
\end{equation}
and there is no pairing in the ``blocking'' or ``breached pairing''
region between these momenta
\cite{LOFF,OurLOFF,Bowers:2001ip,Liu:2002gi,Shovkovy:2003uu,Gubankova:2003uj}
The pairing criterion \eqn{pairing} can be interpreted as
saying that the free energy cost $2 \De$ of breaking a Cooper
pair of two quarks $a$ and $b$ 
is greater than the free energy $2\de\mu$ gained by
emptying the $a$ state and filling the $b$ state (assuming that
$\de\mu$ pushes the energy of the $a$ quark up and the $b$ quark down)
\cite{CFLneutral}.

We now apply these ideas to quark matter.
To impose neutrality, we introduce an electrostatic potential $\mu_e$
coupled to $Q_e$ which is the {\em negative} of the electric charge,
and chemical (color-electrostatic) potentials $\mu_3$ and $\mu_8$
coupled to the diagonal color generators. To take into account the
leading effect of the strange quark mass, we introduce an effective
chemical potential $-M_s^2/(2\mu)$ for the strange quarks. The
splittings of the various pairs are then as given in the middle column
of table \ref{tab:splittings}. We do not discuss the $ru$,
$gd$, and $bs$ quarks that pair in a $3\times 3$ block, because
they do not play a role in defining the boundaries of the
CFL and gCFL phases.

\begin{table}
\newcommand{\st}{\rule[-1.5ex]{0em}{4ex}}
\begin{tabular}{ccc}
\hline
\st quark pair & $\de\mu_{\rm eff}$ 
  & $\de\mu_{\rm eff}$~in electronless CFL \\
\hline
\st $rd$-$gu$& $\frac{1}{2}(\mu_e+\mu_3)$
  & $\mu_e$ \\
\st $rs$-$bu$ & $\frac{1}{2}(\mu_e+\frac{1}{2}\mu_3+\mu_8 - M_s^2/(2\mu))$
  & $\mu_e - M_s^2/(2\mu)$ \\
\st $bd$-$gs$ & $\frac{1}{2}(\frac{1}{2}\mu_3-\mu_8 + M_s^2/(2\mu))$
  & $M_s^2/(2\mu)$  \\
\hline
\end{tabular}
\caption{
Chemical potential splittings ($\mu_{1,2}=\bar \mu \pm \de\mu$)
for the $2\times 2$ pairing blocks. The middle column is for the
general case, with chemical potentials coupled to negative
electric charge ($\mu_e$) and to the diagonal color generators
($\mu_3,\mu_8$). To obtain the last column we fix $\mu_3$ and
$\mu_8$ as functions of $\mu_e$ so that varying $\mu_e$ corresponds to
varying $\mu_{\tilde Q}$.
}
\label{tab:splittings}
\end{table}

\begin{figure}[htb]
\begin{center}
\includegraphics[width=0.7\textwidth]{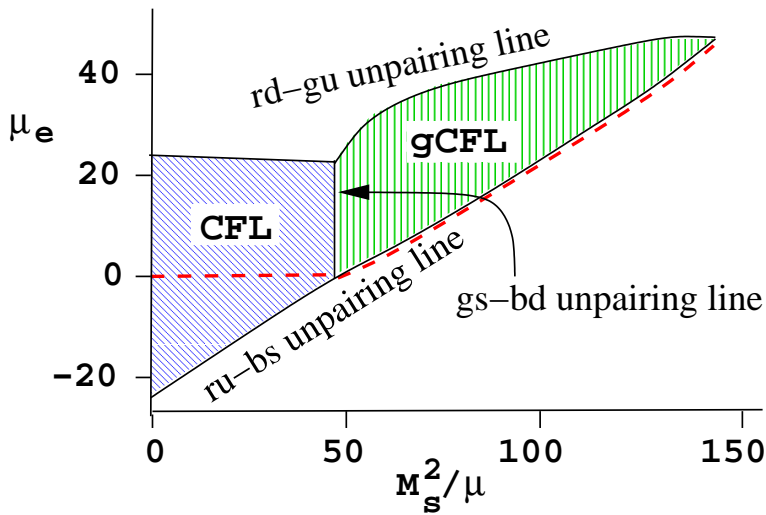}
\end{center}
\vspace{-0.25in}
\caption{Unpairing lines and phases
for the same model as used in Fig.~\ref{fig:gaps}.
If electrons are neglected, then 
the upper and lower curves bound the region of $\mu_e$ 
where CFL or gCFL solutions are found.
Between the curves the quark matter is a $\tilde Q$-insulator.
Taking electrons into account, the correct solution is the dashed
line: in the CFL phase $\mu_e=0$ , and gCFL corresponds
to values of $\mu_e$ below but
very close to the $ru$-$bs$ unpairing line, which is
a $\tilde Q$-conductor because of the ungapped $\tilde Q$-charged
$ru$-$bs$ quasiquarks.
}
\label{fig:mues}
\end{figure}

First consider just the quark matter, with no
electrons (i.e. send the electron mass to infinity).
In this case there is a range of allowed $\mu_e$ at each
$M_s^2/\mu$, because the CFL-paired quark matter
is a $\tilde Q$-insulator, with a free energy that is independent
of $\mu_{\tilde Q}$ as long as all $\tilde Q$-charged modes are gapless.
This means that one can vary $\mu_e$, keeping $\mu_3=\mu_e$
and $\mu_8=\frac{1}{2}(\mu_e-M_s^2/\mu)$, and the free energy 
of the quark matter will not
change, as long the pairing criterion \eqn{pairing} is obeyed
by all quark pairs.
The resultant chemical potential splittings as a function of $\mu_e$
are given in the last column of table \ref{tab:splittings}.
In Fig.~\ref{fig:mues} we show the lines that bound the areas
where the pairing criterion \eqn{pairing} is obeyed, for each
pair in table \ref{tab:splittings}. We see that CFL matter exists in a wedge,
between the $rd$-$gu$ unpairing line and the $rs$-$bu$ unpairing
line, up to a critical value of $M_s^2/\mu$, where the
the $gs$-$bd$ pairs break. From table \ref{tab:splittings} we can see
that the $bd$-$gs$ is vertical because 
the $bd$ and $gs$ are $\tilde Q$-neutral, so
their splitting does not depend on $\mu_e$.
Above the critical $M_s^2/\mu$, the $bd$ and $gs$ are unpaired,
and there is a region in which $rd$-$gu$
and $rs$-$bu$ pairing is maintained.

Now include the electrons. In the CFL region, the system is forced to 
$\mu_e=0$ (dashed line in Fig.~\ref{fig:mues}) \cite{CFLneutral}. 
However,
at the transition point to gCFL,
where the $gs$-$bd$ pairs break, we find that
the neutrality requirement forces us over the line where
$rs$-$bu$ pairs also begin to break.  The result is that 
as $M_s^2/\mu$ increases further, the system
maintains neutrality by staying close to the $rs$-$bu$-unpairing
line, where there is a narrow blocking region in which there
are unpaired $bu$ quarks. Their charge is cancelled by a small
density of electrons.

We see that gCFL quark matter is a conductor of $\tilde Q$
charge, since it has gapless $\tilde Q$-charged quark modes, as
well as electrons. The $rd$ and $gu$ quarks, which are
insensitive to the strange quark mass, remain robustly paired,
and the $\tilde Q$-neutral $bd$ and $gs$ quarks develop a large
blocking region, as the system moves far beyond their unpairing line.
The neutrality  requirement naturally keeps the system close to
the $rs$-$bu$-unpairing line, so these quarks have a very narrow 
blocking region. We can see the effect of this by 
taking $p-\bar\mu \ll \De$ with $\de\mu\approx\De$,
and expanding \Eqn{disprel}:
the dispersion relation for these quarks is almost quadratic,
with a very high density of states at the lowest energies.

We conclude that gCFL quark matter is likely to have very different
transport properties from CFL quark matter. The occurrence of many
gapless fermionic modes will certainly affect cooling, and the
fact that some of them are $\tilde Q$-charged will affect the
conductivity and hence the behavior of magnetic fields. It remains to
be seen what astrophysical signatures flow from these properties,
and this question is being actively studied.
In the rest of these proceedings we will
discuss observable properties of compact stars, however we will
not try to treat the complications of the gCFL phase: we will
assume that the pairing is strong enough
or the strange quark is light enough so that quark matter
always occurs in the CFL phase.

\section{Compact star transport phenomenology}
\label{sec:phenomenology}

The high density and relatively low temperature required to produce
color superconducting quark matter may be attained
in compact stars.
Typical compact stars have masses
close to $1.4 M_\odot$, and are believed to have radii of order 10 km.

Color superconductivity affects the equation of state
at order $(\Delta/\mu)^2$. It also gives mass to
excitations around the ground state: it opens
up a gap at the quark Fermi surface, and makes the gluons
massive. One would therefore expect it to have a profound effect
on transport properties, such as mean free paths,
conductivities and viscosities.
Various observable consequences are under investigation.

\begin{itemize}
\item[\Dash] {\em $r$-mode spindown}.
The $r$-mode is a
bulk flow in a rotating star that, if the
viscosity is low enough, radiates away energy and angular momentum
in the form of gravitational waves.
One can rule out certain models for compact stars
on the grounds that they have such low damping that 
they could not support the high rotation rates observed in pulsars:
$r$-mode spindown would have slowed them down.
Madsen~\cite{MadsenRmode} has shown that
for a compact star made {\em entirely} of quark matter
in the CFL phase, even a gap as small as $\Delta=1$~MeV
is ruled out by observations of millisecond pulsars.
It remains to extend this calculation to the more generic picture of
a quark matter core surrounded by a nuclear mantle.

\item[\Dash] {\em Interfaces and mixed phases.}
These were studied in
Ref.~\cite{ARRW}, and it was found that a mixed
phase only occurs if the surface tension of the interface
is less than about $40~{\rm MeV}/{\rm fm}^2 = 0.2\times (200~{\rm MeV})^3$, 
a fairly small value
compared to the relevant scales $\Lambda_{\rm QCD}\approx 200~{\rm MeV}$,
$\mu\sim 400~{\rm MeV}$. 
A sharp nuclear-quark interface will have an energy-density discontinuity
across it, which will affect gravitational waves emitted in
mergers, and also the $r$-mode spectrum
and the damping forces to which $r$-modes are subject.

\item[\Dash] {\em Crystalline pairing (the ``LOFF'' phase)}.
This is expected to occur when two different types of quark have sufficiently
different Fermi momenta
that BCS pairing cannot occur \cite{OurLOFF}.
This is a candidate for the intermediate  ``non-CFL''
phase of Fig.~\ref{fig:phase},
where the strange quark mass, combined with requirements of weak equilibrium
and charge neutrality, gives each quark flavor a different Fermi momentum.
The phenomenology of the crystalline phase has not yet been worked out,
but recent calculations using Landau-Ginzburg effective theory 
indicate that the favored phase may be a face-centered cubic crystal
 \cite{Bowers:2002xr}, with a reasonably large binding energy.
This raises the interesting possibility of glitches in quark matter stars.

\item[\Dash] {\em Cooling by neutrino emission}.
The cooling rate is
determined by the heat capacity and emissivity, both
of which are sensitive to the spectrum of low-energy excitations,
and hence to color superconductivity.
CFL quark matter, where all modes are gapped, has a much
smaller neutrino emissivity and heat capacity than nuclear matter, and
hence the cooling of a compact star is likely to be dominated by the
nuclear mantle rather than the CFL core 
\cite{Page,Shovkovy-02,Jaikumar:2002vg}.  
Other phases such as 2SC or LOFF give large gaps to only
some of the quarks. Their cooling would proceed quickly, then
slow down suddenly when the temperature fell below
the smallest of the small weak-channel gaps. This behavior should be
observable \cite{Reddy:2003ap}.

\end{itemize}

\section{Mass-radius relationship for compact stars}
\label{sec:massradius}

Although the effects of color superconductivity on the quark matter
equation of state are subdominant, they may have a large effect
on the mass-radius relationship. The reason for this is that
the pressure of quark matter relative to the hadronic vacuum
contains a constant (the ``bag constant'' $B$) that represents
the cost of dismantling the chirally broken and confining
hadronic vacuum,
\begin{equation}
p = (1-c)\frac{3}{4\pi^2}\mu^4 
- \frac{3}{4\pi^2}m_s^2\mu^2 
+ \frac{3}{\pi^2} \Delta^2\mu^2 
- B~.
\label{roughEoS}
\end{equation}
If the bag constant is large enough so that nuclear matter is favored
(or almost favored) over quark matter at $\mu\sim 320~{\rm MeV}$, then the
bag constant and $\mu^4$ terms almost cancel, so if we can fix
the bag constant by other means then the 
strange quark mass $m_s$ and color superconducting
gap $\Delta$ may have a large effect on the equation of state and hence
on the mass-radius relationship of a compact star \cite{Lugones:2002ak}.

\subsection{$M(R)$ at fixed bag constant}

\begin{figure}[htb]
\begin{center}
\includegraphics[width=0.8\textwidth]{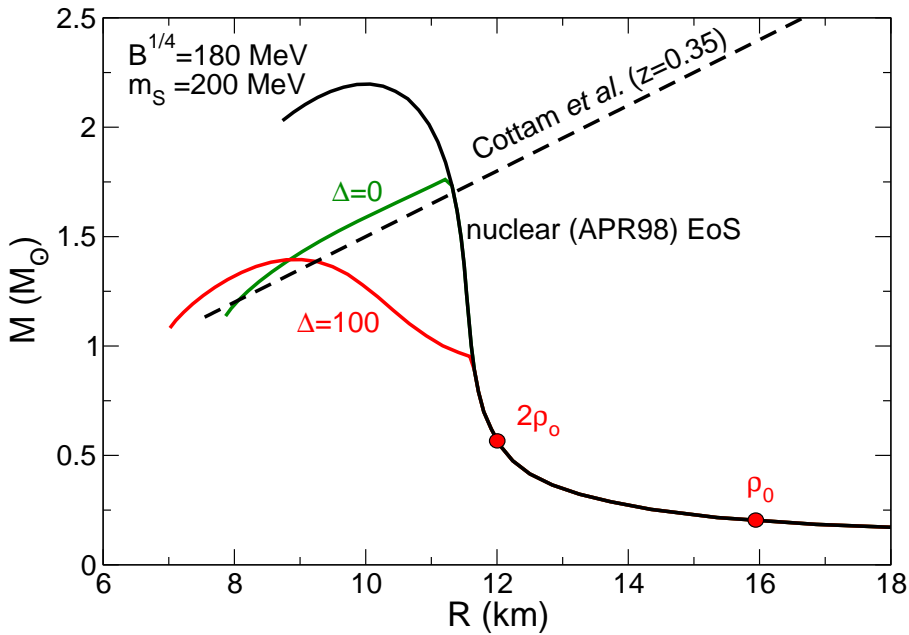}
\end{center}
\caption{Mass-radius relationships for APR98 nuclear matter, competing
with quark matter with fixed bag constant
$B^{1/4}=180~{\rm MeV}$ and $m_s=200$ MeV, either unpaired ($\Delta=0$) 
or CFL color-superconducting ($\Delta=100~{\rm MeV}$).
The dots labeled
$\rho_0$ and $2\rho_0$ on the nuclear matter mass-radius curve
indicate that the central density at these locations correspond to
nuclear and twice nuclear saturation density respectively. }
\label{fig:massradius}
\end{figure}

In Ref.~\cite{AlfordReddy} Sanjay Reddy and I explored the effect of quark
pairing on the $M$-$R$ relationship at fixed values of the bag constant that
are consistent with nuclear phenomenology.  
Fig.~\ref{fig:massradius}
shows the mass-radius curve for the bag model of dense matter, in which
there is competition between a nuclear matter phase
and a quark matter phase.
The nuclear matter was described either by the
APR98 equation of state \cite{APR98}.
The quark matter equation of state was essentially that of 
equation \eqn{roughEoS},
but we included the full (free-quark) correction due to the strange quark mass.
The coefficient of the $\Delta^2\mu^2$ term is the one appropriate to
CFL color superconductivity involving all nine colors/flavors of the
quarks. We used physically reasonable values of the bag constant 
$B^{1/4}=180$ MeV ($B=137~{\rm MeV}\!/{\rm fm}^3$) and
strange quark mass $m_s=200$ MeV. 
We set the parameter $c$ to zero: its effects will be discussed below.

Curves for unpaired ($\Delta=0$) and color-superconducting ($\Delta=100~{\rm MeV}$)
quark matter are shown. At these  values
the stars are typically ``hybrid'',
containing both quark matter and nuclear matter.  The solid lines in
Fig.~\ref{fig:massradius} correspond to stars that either have no QM
at all, or a sharp transition between NM and QM: the core is made of
quark matter, which is the favored phase at high pressure, and at some
radius there is a transition to nuclear matter, which is favored at
low pressure.  The transition pressure is sensitive to $\Delta$, for
reasons discussed earlier. The dashed lines are for stars that contain
a mixed NM-QM phase.  In all cases we see that light, large stars
consist entirely of nuclear matter. When the star becomes heavy
enough, the central pressure rises to a level where QM
occurs in the core. As can be seen
from the figure the transition density is very sensitive to $\Delta$.
The line labeled ``Cottam {\it et al.}''
indicates the constraint obtained by recent measurements of the
redshift on three spectral lines from EXO0748-676 \cite{Cottam}.

\subsection{$M(R)$ with non-free quarks: quark matter mimicking
nuclear matter}

In Ref.~\cite{AlfordReddy} we kept the bag constant fixed,
assuming that it could be fixed by other observations, and 
we treated the quark matter as free quarks with a pairing
energy. It is interesting to see what happens when we relax 
these assumptions, since the bag constant is not easily
measured, and even after taking pairing into account 
we expect remaining QCD interactions between the quarks in the Fermi sea.

To allow for effects of quark interactions beyond Cooper pairing,
we follow the parameterization of Fraga et.~al.~\cite{Fraga},
who find that the
${\cal O}(\alpha_s^2)$ pressure for three unpaired flavors over
the relevant range of
$\mu$ is well-described by a bag-model-inspired form given by
\begin{equation}
P_{\alpha_s^2}(\mu) = 
\displaystyle \frac{3}{4\pi^2}~a_{\rm eff}~\mu^4 - B_{\rm eff}\, , \qquad
a_{\rm eff} \equiv  1-c\ .
\label{peff}
\end{equation}
They find $a_{\rm eff}\approx 0.63$ ($c\approx 0.37$),
but  admit that at the density of interest for compact
star physics the QCD coupling is strong, and there
there is no reason to expect the leading order calculation to be
accurate. We therefore take their result as indicating that it
is reasonable to treat $c$ as an additional parameter in the quark
matter equation of state, as shown in \Eqn{roughEoS},
and we proceed to study its effects
on the mass-radius relationship of compact stars.

To see how closely quark matter can mimic nuclear matter, we will
not treat the bag constant as fixed, but
tune it to keep the physics as constant as possible.
Thus when we compare, say, $\Delta=0$ (non-color-superconducting)
quark matter with CFL ($\Delta=50$) quark matter, we set the bag constant
in each case so that the transition from nuclear to
quark matter occurs at a given density of nuclear matter.
This effectively ``subtracts out'' the part of any variation
in $\Delta$ that simply corresponds to a renormalization
of the bag constant, which is in any case very poorly known.

\begin{figure}[htb]
\begin{center}
\includegraphics[width=0.7\textwidth,angle=-90]{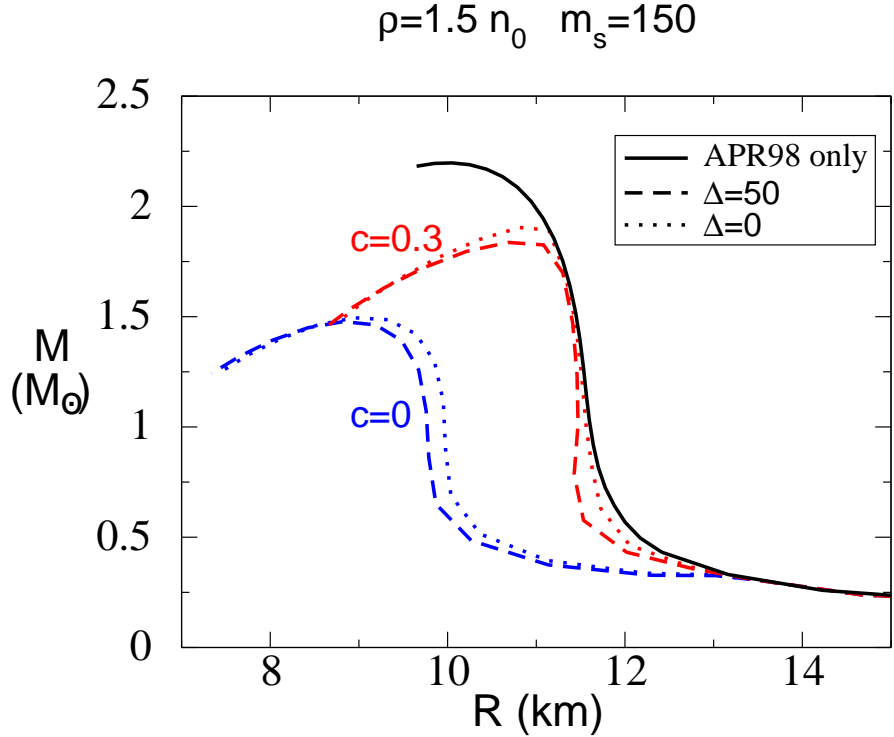}
\end{center}
\caption{$M(R)$ relationship for APR98 nuclear
matter with various quark matter equations of state.
The strange quark is light, and the bag constant is tuned so that
the nuclear matter to quark matter transition occurs at 1.5 times
nuclear saturation density. Dotted lines are unpaired quark matter,
dashed lines are CFL with gap of $50~{\rm MeV}$. 
Note how the curve for CFL quark matter with perturbative correction 
but no color superconductivity ($c=0.3$, $\Delta=0$; red dotted)
closely follows the pure nuclear curve up to $M\approx 1.9 M_\odot$.
}
\label{fig:rho1.5}
\end{figure}

We explore the effect of a color superconducting gap $\Delta$
and perturbative correction $c$ on the mass-radius relationship.
We fix the bag constant by requiring that that the 
nuclear to quark matter phase
transition occur at nuclear matter baryon density 
$\rho=1.5 n_{\rm sat}$. The resultant $M(R)$ curves
are shown in Fig.~\ref{fig:rho1.5}.
The noticeable features of the plots are
\begin{enumerate}
\item {\em Increasing $c$ makes the stars smaller and lighter}.\\
In our parameterization the stars resulting from
quark matter equations of state without perturbative correction
($c=0$, blue lines) are smaller and lighter.

\item {\em Color superconductivity acts like a change in the bag constant.}\\
In Ref.~\cite{AlfordReddy} we showed that at fixed bag constant,
color superconductivity has a strong effect on the mass-radius
relationship of compact stars. Here, by comparing the
dashed lines with the dotted lines in Fig.~\ref{fig:rho1.5},
we see that it is difficult to distinguish the effect of
color superconductivity from a change in the bag constant.
In Fig.~\ref{fig:rho1.5}, as we vary parameters $c$ and $\Delta$
of the quark matter equation of state, the bag constant 
is tuned to maintain a constant value of the nuclear density
at the transition to quark matter, and in this situation
color superconductivity only makes a small difference to the
mass-radius relationship.
\item {\em Quark matter with $c\approx 0.3$ looks 
 just like APR98 nuclear matter.}\\
The stars with $c=0.3$ have mass-radius relationships that are very similar to
the pure nuclear APR98 matter.  In fact, for the case where there are
perturbative corrections but no color superconductivity
the equations of state ($p(\mu)$) are so similar that our program found
a series of phase transitions back and forth between CFL and APR98
up to $\mu=546~{\rm MeV}$ (baryon density $\rho=5.4 n_{\rm sat}$). 
This is why the $c=0.3$ red dotted curve lies almost
exactly on top of the solid black (APR98) curve, even though there
was a phase transition from APR98 to CFL at $\rho=1.5 n_{\rm sat}$
(which is first attained when the APR98 star reaches a mass of
$0.315~M_\odot$, $R=13.3~{\rm km}$). 
\end{enumerate}

\subsection{$M(R)$ measurements and quark matter}

We can now ask what significance mass and radius measurements
will have for the presence of quark matter, and particularly
color-superconducting quark matter, in compact stars.
\begin{itemize}
\item {\em What would rule out quark matter?}\\
From Fig.~\ref{fig:rho1.5} we see that an observed mass $M \gtrsim 2~M_\odot$
would be inconsistent with the star containing quark matter obeying
the equation of state that we have studied here.
However, we emphasize that by introducing the parameter $c$
and setting it to a reasonable value $c\approx 0.3$ we have increased
the mass range for hybrid stars, moving the upper limit from its
old value around $1.6~M_\odot$ up to about $1.9~M_\odot$.
\item {\em What would indicate the presence of quark matter?} \\
This is difficult. Regions of $M$-$R$ space that cannot be reached by
any nuclear matter equation of state also cannot be reached by hybrid
NM-QM equations of state. It is clear from Figs.~\ref{fig:rho1.5}
and \ref{fig:massradius} that hybrid stars are smaller
than Walecka or APR98 nuclear matter stars,
with radii of around $10~{\rm km}$ at $M\approx 1.4~M_\odot$.
But there are many other suggested nuclear equations
of state, and the flattening of the $M(R)$ curve that appears in our
plots to be characteristic of quark matter may easily be mimicked
by kaon condensation in nuclear matter \cite{Lattimer:2000nx}.


Obviously the region of pure quark matter objects which lie
at very low mass and radius (``strangelets'') is not attainable
by nuclear matter, but the existence of such objects, unlike
that of compact stars, remains a matter of speculation.

\item {\em What would indicate the presence
of \underline{color superconducting} quark matter?} \\
This is more difficult.
Even if we found an $M(R)$ characteristic of quark matter,
we would need an independent determination of the bag constant to 
claim that it was color-superconducting.
\end{itemize}

\begin{center}{\bf Acknowledgments}\end{center}
I thank the organizers of ``Finite density QCD
at Nara'' and Confinement 2003. The work reported in Section 
\ref{sec:massradius} was
performed in collaboration with Sanjay Reddy, and was supported
by the UK PPARC and by the U.S. Department of Energy under grant number
DE-FG02-91ER40628. The work reported in Section \ref{sec:gCFL}
was performed in collaboration with C. Kouvaris and K. Rajagopal,
and was supported in part by DOE grants 
DE-FG02-91ER40628 and DF-FC02-94ER40818.


\begin{thebibliography}{99}

\bibitem{Reviews}
K.~Rajagopal and F.~Wilczek,
hep-ph/0011333.
%
M.~G.~Alford,
Ann.\ Rev.\ Nucl.\ Part.\ Sci.\  {\bf 51} (2001) 131
[hep-ph/0102047].
%
T.~Schaefer,
hep-ph/0304281.
%
D.~H.~Rischke,
nucl-th/0305030.
%
D.~K.~Hong,
Acta Phys.\ Polon.\ B {\bf 32} (2001) 1253
[hep-ph/0101025].
%

\bibitem{ARW3}
M.~Alford, K.~Rajagopal and F.~Wilczek, Nucl. Phys. {\bf B537}, 443 (1999) 
[hep-ph/9804403].

\bibitem{gCFL}
M. Alford, C. Kouvaris, K. Rajagopal, hep-ph/0311286.

\bibitem{OurLOFF}
M.~Alford, J.~Bowers and K.~Rajagopal,
Phys.\ Rev.\ D {\bf 63}, 074016 (2001)
[hep-ph/0008208].


\bibitem{oneflav}
M.~Alford, J.~Bowers, J.~Cheyne and G.~Cowan,
Phys.\ Rev.\ D {\bf 67}, 054018 (2003)
[hep-ph/0210106].
%
M.~Buballa, J.~Hosek and M.~Oertel,
hep-ph/0204275.
%
T.~Sch\"afer,
Phys. Rev. {\bf D62}, 094007 (2000). 

\bibitem{IwaIwa}
M.~Iwasaki, T.~Iwado, Phys. Lett. {\bf B350}, 163 (1995);
M.~Iwasaki, Prog. Theor. Phys. Suppl. {\bf 120}, 187 (1995)


\bibitem{oldcolorSC}
B.~Barrois, Nucl.~Phys.~{\bf B129} (1977) 390;
``Nonperturbative effects in dense quark matter'',
Cal Tech PhD thesis, UMI 79-04847-mc (1979).
S.~Frautschi, Proceedings of workshop on hadronic matter at extreme density,
Erice 1978, pp 19-27.
D.~Bailin and A.~Love, Phys. Rept. {\bf 107} (1984) 325,
and references therein.

\bibitem{newcolorSC}
M.~Alford, K.~Rajagopal and F.~Wilczek,
Phys.\ Lett.\  {\bf B422}, 247 (1998)
[hep-ph/9711395].
R.~Rapp, T.~Sch\"afer,
E.~V.~Shuryak and M.~Velkovsky, Phys. Rev. Lett. {\bf 81}, 53 (1998) 
[hep-ph/9711396].

\bibitem{BCS}
J.~Bardeen, L.~Cooper, J.~Schrieffer, Phys. Rev. {\bf 106}, 162 (1957);
Phys. Rev. {\bf 108}, 1175 (1957)


\bibitem{SW-cont}
T.~Sch\"afer, F.~Wilczek, Phys. Rev. Lett. {\bf 82}, 3956 (1999)
[hep-ph/9903503].

\bibitem{HsuCFL}
N.~Evans, J.~Hormuzdiar, S.~Hsu, M.~Schwetz:
  Nucl. Phys. {\bf B581}, 391 (2000)
[hep-ph/9910313].

\bibitem{3flavpert}
T.~Sch\"afer,
Nucl.\ Phys.\  {\bf B575}, 269 (2000)
[hep-ph/9909574].
I.~Shovkovy, L.~Wijewardhana, Phys. Lett. {\bf B470}, 189 (1999).

\bibitem{PisarskiCFL} 
R.~D.~Pisarski and D.~H.~Rischke,
``Why color-flavor locking is just like chiral symmetry breaking''.
To be published in, {\it Proceedings of the Judah Eisenberg
  Memorial Symposium, ``Nuclear Matter, Hot and Cold''},
  Tel Aviv, April 14 - 16, 1999 [nucl-th/9907094].

\bibitem{BaymIida}
K.~Iida and G.~Baym,
Phys.\ Rev.\ D {\bf 63}, 074018 (2001)
[hep-ph/0011229].

\bibitem{AR-02}
M.~Alford and K.~Rajagopal,
JHEP {\bf 0206} (2002) 031
[hep-ph/0204001].


\bibitem{Steiner:2002gx}
A.~W.~Steiner, S.~Reddy and M.~Prakash,
Phys.\ Rev.\ D {\bf 66}, 094007 (2002)
[hep-ph/0205201].


\bibitem{ARRW}
M.~G.~Alford, K.~Rajagopal, S.~Reddy and F.~Wilczek,
hep-ph/0105009.


\bibitem{Schaefer:2002gf}
T.~Sch\"afer,
Phys.\ Rev.\ D {\bf 65} (2002) 094033.

\bibitem{BedaqueSchaefer}
P.~F.~Bedaque and T.~Sch\"afer,
Nucl.\ Phys.\ A {\bf 697} (2002) 802
[hep-ph/0105150].



\bibitem{ABR2+1}
M.~Alford, J.~Berges and K.~Rajagopal,
Nucl.\ Phys.\  {\bf B558}, 219 (1999)
[hep-ph/9903502].

\bibitem{LOFF}
A. I. Larkin and Yu. N. Ovchinnikov, Zh. Eksp. Teor. Fiz. {\bf 47}, 1136
(1964) [Sov. Phys. JETP {\bf 20}, 762 (1965)];
P.~Fulde and R.~A.~Ferrell, Phys.\ Rev.\ {\bf 135}, A550 (1964);
S.~Takada and T.~Izuyama, Prog.\  Theor.\ Phys.\ {\bf 41}, 635 (1969);

\bibitem{Bowers:2002xr}
J.~A.~Bowers and K.~Rajagopal,
hep-ph/0204079.

\bibitem{Bowers:2001ip}
J.~A.~Bowers, J.~Kundu, K.~Rajagopal and E.~Shuster,
Phys.\ Rev.\ D {\bf 64}, 014024 (2001).

\bibitem{Liu:2002gi}
W.~V.~Liu and F.~Wilczek,
Phys.\ Rev.\ Lett.\  {\bf 90}, 047002 (2003)
[arXiv:cond-mat/0208052].



\bibitem{Shovkovy:2003uu}
I.~Shovkovy and M.~Huang,
Phys.\ Lett.\ B {\bf 564}, 205 (2003);
M.~Huang and I.~Shovkovy,
arXiv:hep-ph/0307273.

\bibitem{Gubankova:2003uj}
E.~Gubankova, W.~V.~Liu and F.~Wilczek,
Phys.\ Rev.\ Lett.\  {\bf 91}, 032001 (2003).




\bibitem{CFLneutral}
K.~Rajagopal and F.~Wilczek,
Phys.\ Rev.\ Lett.\  {\bf 86}, 3492 (2001)
[hep-ph/0012039].

\bibitem{MadsenRmode}
J.~Madsen,  Phys. Rev. Lett. {\bf 85}, 10 (2000) [astro-ph/9912418].

\bibitem{Page}
D.~Page, M.~Prakash, J.~Lattimer, A.~Steiner, 
 Phys. Rev. Lett. {\bf 85} (2000) 2048
[hep-ph/0005094].

\bibitem{Shovkovy-02}
I.~Shovkovy, P.~Ellis
[astro-ph/0207346].

\bibitem{Jaikumar:2002vg}
P.~Jaikumar, M.~Prakash and T.~Schafer,
Phys.\ Rev.\ D {\bf 66}, 063003 (2002)
[astro-ph/0203088].

\bibitem{Reddy:2003ap}
S.~Reddy, M.~Sadzikowski and M.~Tachibana,
nucl-th/0306015.

\bibitem{Lugones:2002ak}
G.~Lugones and J.~E.~Horvath,
hep-ph/0211070.

\bibitem{AlfordReddy}
M.~Alford and S.~Reddy,
Phys.\ Rev.\ D {\bf 67}, 074024 (2003)
[nucl-th/0211046].


\bibitem{APR98}
A. Akmal, V.R. Pandharipande, D.G. Ravenhall,
Phys.Rev. C58 1804 (1998)
[nucl-th/9804027].

\bibitem{Cottam}
J.~Cottam, F.~Paerels, M.~Mendez, Nature {\bf 420}, 51 (2002).

\bibitem{Fraga}
E.~S.~Fraga, R.~D.~Pisarski and J.~Schaffner-Bielich,
Phys.\ Rev. {\bf D63}, 121702 (2001)
[hep-ph/0101143].

\bibitem{Lattimer:2000nx}
J.~M.~Lattimer and M.~Prakash,
Astrophys.\ J.\  {\bf 550}, 426 (2001)
[astro-ph/0002232].

\end{thebibliography}
\end{document}